\documentclass[sigconf]{acmart}
\usepackage{multirow}
\usepackage{color}
\usepackage{enumitem}

\usepackage{amssymb}
\usepackage{pifont}
\usepackage{fontawesome}

\AtBeginDocument{%
  }

\setcopyright{acmlicensed}
\copyrightyear{2024}
\acmYear{2024}
\acmDOI{XXXXXXX.XXXXXXX}

\acmConference[CIKM '26]{Make sure to enter the correct
  conference title from your rights confirmation email}{November 7-11, 2026}{Rome, ITALY}
\acmISBN{978-1-4503-XXXX-X/18/06}

\begin{document}

\title[HiGR: Hierarchical Generative Slate Recommendation]
{HiGR: Industrial-Scale Hierarchical Generative Slate Recommendation Framework in Tencent}



 

\author{Yunsheng Pang}
\authornote{Both authors contributed equally to this research.}
\affiliation{%
  \institution{Platform and Content Group, Tencent}
  \country{}
}
\email{yspang@tencent.com}

\author{Zijian Liu}
\authornotemark[1]
\affiliation{%
  \institution{Platform and Content Group, Tencent}
  \country{}
}
\email{kevinsliu@tencent.com}

\author{Yudong Li}
\affiliation{%
  \institution{Platform and Content Group, Tencent}
  \country{}
}
\authornote{Corresponding author.}
\email{elsonli@tencent.com}

\author{Shaojie Zhu}
\affiliation{%
  \institution{Platform and Content Group, Tencent}
  \country{}
}

\author{Zijian Luo}
\affiliation{%
  \institution{Sun Yat-sen University}
  \country{}
}

\author{Chenyun Yu}
\affiliation{%
  \institution{Sun Yat-sen University}
  \country{}
}

\author{Sikai Wu}
\affiliation{%
  \institution{Platform and Content Group, Tencent}
  \country{}
}

\author{Shichen Shen}
\affiliation{%
  \institution{Platform and Content Group, Tencent}
  \country{}
}

\author{Cong Xu}
\affiliation{%
  \institution{Platform and Content Group, Tencent}
  \country{}
}

\author{Bin Wang}
\affiliation{%
  \institution{Platform and Content Group, Tencent}
  \country{}
}

\author{Kai Jiang}
\affiliation{%
  \institution{Platform and Content Group, Tencent}
  \country{}
}


\author{Chengxiang Zhuo}
\affiliation{%
  \institution{Platform and Content Group, Tencent}
  \country{}
}

\author{Zang Li}
\affiliation{%
  \institution{Platform and Content Group, Tencent}
  \country{}
}

\renewcommand{\shortauthors}{Y. Pang et al.}

\begin{abstract}

Slate recommendation, which presents users with a ranked item list in a single display, is ubiquitous across mainstream online platforms. 
While recent generative recommendation methods have shown strong potential in modeling item sequences with semantic IDs, directly applying them to industrial-scale slate recommendation faces a fundamental disconnect: entangled SID spaces confound high-level list planning, fine-grained autoregressive decoding over long sequences limits semantic planning efficiency, and token-level objectives misalign with holistic slate quality.
In this paper, we propose HiGR, an industrial-scale hierarchical generative framework for slate recommendation that bridges this disconnect through a co-designed pipeline. First, HiGR learns structured SIDs via a Prefix-Contrastive Residual Quantized VAE (PCRQ-VAE). By enforcing high-level prefixes to capture shared semantics, PCRQ-VAE creates a controllable discrete space that acts as a prerequisite for efficient planning. Leveraging this structured space, our Hierarchical Slate Decoder (HSD) shifts autoregressive modeling from entangled token-level decoding to coarse-grained preference embeddings. This design significantly reduces inference latency while allowing explicit global slate structure planning. Finally, this stable planning space enables an ORPO-based listwise alignment mechanism to optimize triple-objective implicit feedback—ranking fidelity, genuine user interest, and diversity. Extensive offline experiments show that HiGR outperforms state-of-the-art baselines by over 10\% in offline recommendation quality while achieving a $5\times$ inference speedup. Online A/B tests on Tencent platforms further improve watch time by 1.22\% and video plays by 1.73\%. HiGR has been deployed on multiple Tencent platform surfaces (e.g., news, videos, and web novels), serving hundreds of millions of users and proving its industrial-scale applicability.


\end{abstract}





\maketitle

\section{Introduction}

Personalized recommendation systems play a central role in large-scale online platforms, such as video feeds, news feeds, e-commerce, and homepage recommendation services~\cite{wu2022feedrec,chen2024multi,zhao2015commerce,liu2025onerecthinkintextreasoninggenerative}. In many of these scenarios users are shown an ordered list of items in a single display—a setting commonly known as slate recommendation—where quality depends not only on individual item relevance but also on the inter-item structure of the list (ordering, compatibility, and diversity).

Most industrial recommendation systems still follow a discriminative two-stage paradigm. They first score candidate items independently with point-wise or pair-wise ranking models~\cite{lei2017alternating}, and then assemble the final slate through greedy selection, heuristic rules, or reranking modules. Optimizing items in isolation limits the ability to capture inter-item dependencies and positional effects, often yielding slates that are locally reasonable but globally suboptimal.

Recent advances in generative recommendation provide a promising alternative. By formulating recommendation as a sequence generation problem, generative models can directly produce an ordered slate and naturally model dependencies among items~\cite{deffayet2023generative, deng2025onerec}. In particular, semantic-ID-based methods tokenize items into compact discrete codes through vector quantization, making it possible to generate items in a language-model-like manner without relying on an extremely large item vocabulary~\cite{rajput2023recommender, wang2024learnable}. This generative formulation is conceptually appealing for slate recommendation, because it shifts the task from independently selecting items to directly generating a structured list.

However, directly applying existing generative recommendation methods to slate recommendation exposes a fundamental disconnect between item-level generation and holistic slate optimization in three aspects. First, generative recommendation relies on high-quality item tokenization, but existing semantic ID methods often produce entangled code spaces, where inconsistent SID prefixes weaken controllability and hinder slate-level relevance and diversity guidance. Second, slate generation with semantic IDs remains inefficient: an $M$-item slate requires decoding $M \times D$ SID tokens (with $D$ SIDs per item), while full-sequence autoregressive beam search introduces substantial latency and entangles intra-item semantics with inter-item transitions, obscuring global slate structure. Third, token-level likelihood training does not directly optimize user-perceived slate quality. Real-world recommendation platforms care about multiple list-level objectives, such as ranking fidelity, genuine user interest, and intra-list diversity, which cannot be fully captured by next-token prediction alone.

To address these challenges, we propose \textbf{HiGR}, an industrial-scale hierarchical generative framework for slate recommendation. HiGR builds a unified pipeline from structured item tokenization, to efficient hierarchical slate generation, and finally to slate-level preference alignment. Specifically, we first design a Prefix-Contrastive Residual Quantized VAE (PCRQ-VAE) to learn semantically structured item IDs. By introducing prefix-level contrastive alignment into residual quantization, PCRQ-VAE encourages semantically or collaboratively similar items to share high-level SID prefixes, while preserving the final code layer for fine-grained item discrimination. This provides a more controllable discrete space for subsequent slate generation. Based on the structured SID space, we further propose a Hierarchical Slate Decoder (HSD) for efficient coarse-to-fine slate generation. Instead of decoding the entire $M \times D$ SID sequence autoregressively, HSD first plans the slate at the preference-embedding level, where each embedding represents the intended semantics of one slate position. A lightweight item generator then decodes the corresponding SID sequence conditioned on each preference embedding. By shifting global autoregressive modeling from fine-grained SID tokens to coarse-grained preference embeddings and restricting beam search to short item-level decoding, HSD substantially reduces inference latency while preserving high-quality slate generation. Finally, to bridge the gap between token-level generation and real user experience, we introduce a listwise multi-objective preference alignment mechanism. Instead of relying on item-level rewards, we construct slate-level preference pairs from implicit user feedback and optimize the model with Odds Ratio Preference Optimization (ORPO). The preference pairs are designed to reflect three complementary objectives: ranking fidelity, genuine user interest, and slate diversity. In this way, HiGR directly aligns the generated slate with holistic user preferences and practical business metrics.

Our main contributions are summarized as follows:
\begin{itemize}[leftmargin=*]
\item We propose HiGR, an end-to-end generative slate recommendation framework that significantly boosts both inference efficiency and recommendation performance, enabling its deployment on large-scale online platforms.
\item We develop three tightly coupled components, PCRQ-VAE, HSD, and ORPO-based preference alignment, forming a unified pipeline from structured SID tokenization to efficient slate generation and listwise quality optimization.
\item Extensive offline experiments and online A/B tests demonstrate the effectiveness of HiGR. It consistently outperforms state-of-the-art baselines in recommendation quality and achieves significant inference acceleration. HiGR has also been deployed on multiple business scenarios on Tencent platforms, serving hundreds of millions of users and bringing consistent online improvements.
\end{itemize}

\section{Related Work}
\subsection{Slate Recommendation}
Standard pointwise recommendation methods 
\cite{guo2017deepfm,kang2018self,cheng2016wide,sun2019bert4rec,covington2016deep,hou2025generating,qiu2025one} 
score user-item pairs independently. 
Although effective for individual feedback prediction such as CTR, they largely ignore inter-item dependencies and positional effects. 
Slate recommendation instead optimizes the whole list to capture global context and item interactions. 
Representative methods include ListCVAE \cite{jiang2018beyond}, which models list distributions with conditional VAEs; 
GFN4Rec \cite{liu2023generative}, which formulates list generation as a reward-proportional flow; 
and DMSG \cite{tomasi2024diffusion}, which adopts conditional diffusion models for coherent and diverse slate generation. 
However, these methods often rely on reranking pipelines or complex objectives, causing cascading errors or training instability. 
In contrast, we formulate slate construction as end-to-end autoregressive generation, enabling scalable global planning for industrial deployment.

\subsection{Generative Recommendation}
Inspired by LLMs, recommendation systems are shifting from discriminative scoring to generative modeling \cite{hu2025ids,dai2025onepiecebringingcontextengineering,zhang2025gpr,zhang2025slow,zhai2024actions,han2025mtgr,zhang2025onetrans}, where recommendation is formulated as a sequence generation task. Since directly using items as tokens leads to a huge vocabulary and poor cold-start generalization \cite{chen2024enhancing}, Semantic IDs from vector quantization have been widely adopted to represent items with compact discrete codes. Representative methods include TIGER \cite{rajput2023recommender}, which uses RQ-VAE for recursive item tokenization, LETTER \cite{wang2024learnable}, which learns codebooks end-to-end, and OneRec \cite{deng2025onerec}, which constructs hierarchical IDs via iterative K-means. However, these methods mainly align item features before or after quantization, leaving the internal SID hierarchy insufficiently structured. In contrast, our Prefix-Contrastive RQ-VAE enforces prefix-level alignment during quantization, making high-level SID prefixes more semantic and controllable for hierarchical slate generation.

Preference alignment bridges next-token prediction and business metrics. PrefRec \cite{xue2023prefrec} uses an RLHF-style reward model, while DPO-based methods \cite{rafailov2023direct,deng2025onerec,zhou2025onerec} offer more stable optimization for generative recommenders. However, they often require reference models or rely on generic/item-level objectives, failing to capture the multi-dimensional nature of slate quality. We adopt reference-free ORPO and design triple-objective slate-level preferences to jointly improve ranking fidelity, genuine interest, and diversity.

\begin{figure*}
  \centering
  \includegraphics[width=0.95\textwidth,height=0.42\textheight,keepaspectratio=false]{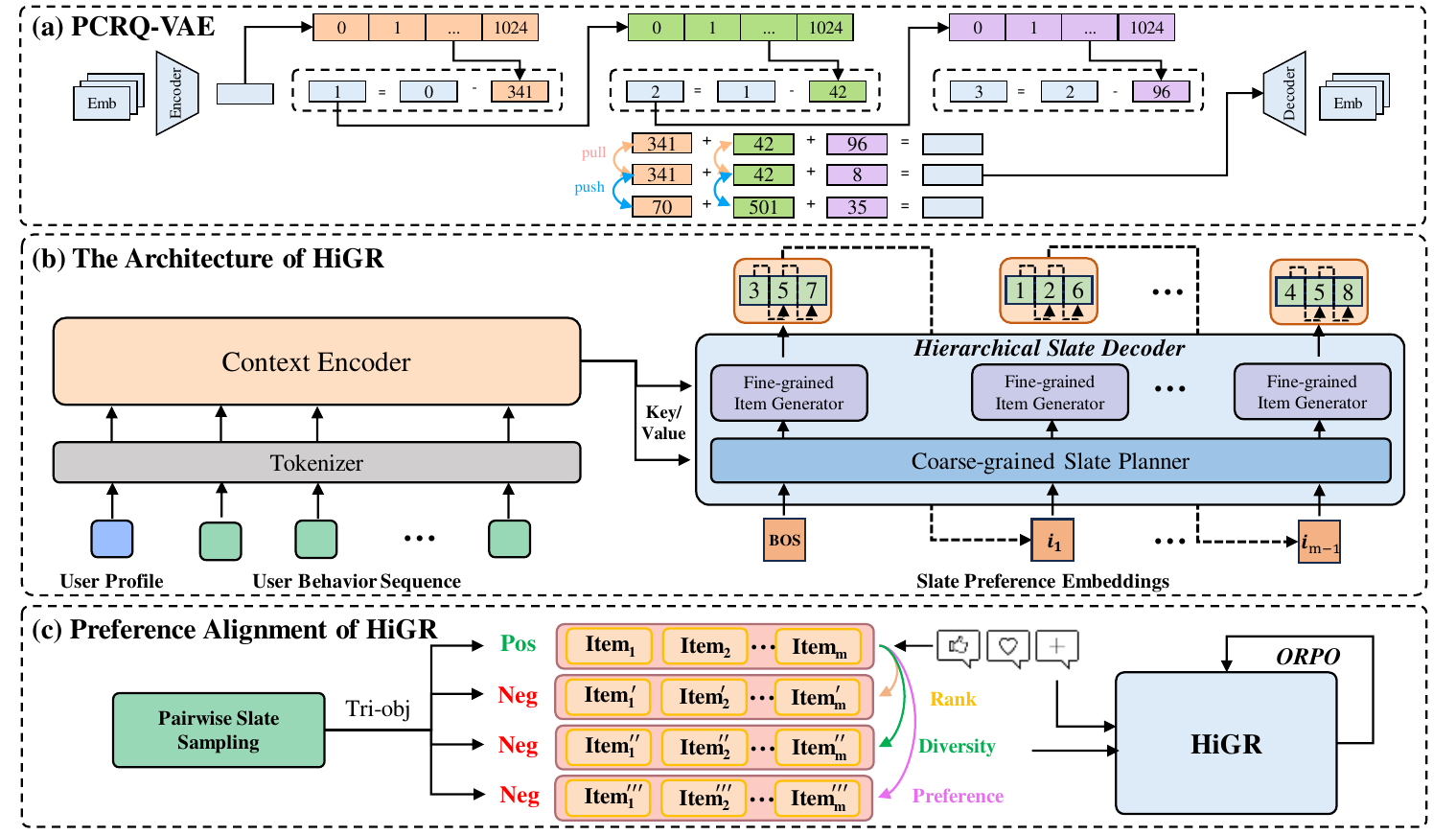}
  \vspace{-0.2cm}
  \caption{The overall framework of HiGR, which contains (a) PCRQ-VAE for semantic tokenization, (b) Hierarchical Slate Decoder for coarse-to-fine generation, and (c) ORPO-based Preference Alignment for slate quality optimization.}
  \label{HiGR} 
 \vspace{-0.3cm}
\end{figure*}


\section{The HiGR Framework}

\subsection{Framework Overview}
Slate recommendation aims to generate an ordered item list that optimizes user feedback at the list level. 
Let $\mathcal{U}$ and $\mathcal{V}$ denote the user and item sets. 
For each $u \in \mathcal{U}$ with chronological interaction history $S^u=\{v_1^u,\cdots,v_n^u\}$, the model generates an $M$-item slate:
\begin{equation}
    O^u = \mathcal{F}_{\theta}(u, S^u),
\end{equation}
where $\mathcal{F}_{\theta}$ denotes the recommendation model and $O^u=\{\hat{v}_1^u,\cdots,\hat{v}_M^u\}$ is optimized for holistic slate quality rather than item-wise quality.

In this paper, we propose HiGR, a hierarchical generative framework tailored for high-quality and efficient slate recommendation.
As shown in Figure \ref{HiGR}, HiGR explicitly considers the list-level nature of slate recommendation from three coupled perspectives: First, Prefix-Contrastive RQ-VAE learns hierarchical SIDs whose prefixes encode item-level semantic and collaborative similarity, providing structured discrete tokens for slate generation. 
Second, the Hierarchical Slate Decoder generates a slate by first planning list-level preferences and then decoding item-level SIDs. 
Third, listwise preference alignment further optimizes the generated slate as a whole with multi-objective implicit feedback. 
These components form a unified pipeline from slate-aware item tokenization to efficient slate generation and final slate-level alignment.

\subsection{Prefix-Contrastive RQ-VAE(PCRQ-VAE)} \label{SID}
Semantic IDs (SIDs) are essential for generative slate recommendation, as they convert a slate into discrete token sequences. 
However, standard RQ-based tokenization often produces entangled and sparse ID spaces, where SID prefixes fail to reliably reflect item semantics or collaborative relations. 
This weakens generation controllability and makes it difficult to guide slate-level relevance and diversity through discrete prefixes. To address this issue, we propose Prefix-Contrastive RQ-VAE (PCRQ-VAE), which enhances residual quantization with global quantization and prefix-level contrastive alignment. 
The former stabilizes hierarchical SID learning by preserving informative residuals across codebook layers, while the latter injects semantic and collaborative signals into SID prefixes. 
As a result, similar items are encouraged to share high-level prefixes, whereas the final codebook layer remains available for fine-grained item discrimination.

\begin{figure}
  \centering
  \includegraphics[width=0.45\textwidth]{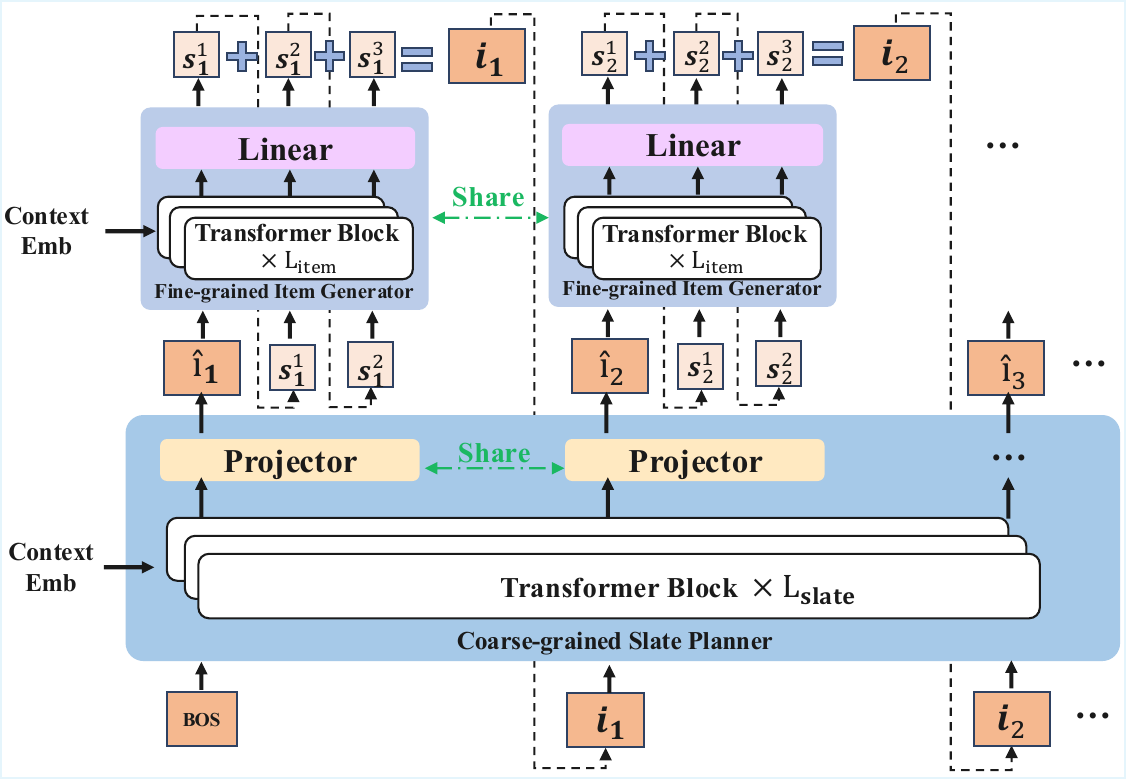}
  \vspace{-0.3cm}
  \caption{The overall architecture of HSD.}
  \label{HSD} 
  \vspace{-0.5cm}
\end{figure}

\subsubsection{Global Quantization for Residual Preservation.} We adopt a standard RQ-VAE backbone to tokenize each item embedding $x$ into a hierarchical SID sequence. The encoder first maps $x$ into a latent representation $z=\mathcal{E}(x)$, which is recursively quantized by $D$ residual codebooks, producing SIDs $\{s^d\}_{d=1}^{D}$ and the aggregated quantized representation $\hat{z}=\sum_{d=1}^{D} e_{s_d}^{d}$. The decoder then reconstructs the input embedding as $\hat{x}=\mathcal{D}(\hat{z})$. Standard RQ-VAE commonly optimizes quantization at each residual layer independently. However, this layer-wise constraint tends to pull each residual $r_{d-1}$ too close to its selected codeword, causing residual vanishing in deeper codebooks and limiting their ability to capture fine-grained semantics. To address this issue, we introduce a global quantization objective that directly aligns the aggregated quantized representation $\hat{z}$ with the original latent embedding $z$: \begin{equation} \mathcal{L}_{\mathrm{global\_quan}} = \|\hat{z}-\mathrm{sg}(z)\|_2^2 + \eta \|z-\mathrm{sg}(\hat{z})\|_2^2 , \end{equation} where $\mathrm{sg}(\cdot)$ denotes the stop-gradient operation and $\eta$ controls the commitment strength. By optimizing quantization at the global latent level, this objective preserves informative residuals across codebook layers and stabilizes hierarchical SID learning.

\subsubsection{Prefix-level Contrastive Alignment.} Although global quantization stabilizes hierarchical SID learning, reconstruction-based objectives alone cannot explicitly align the relational structure among items. To inject collaborative and semantic signals into the SID space, we introduce prefix-level contrastive alignment over the first $D-1$ codebook layers. For each anchor item $a$, we construct a positive item $p$ from semantic neighbors or high co-occurrence signals, while treating other in-batch items as negatives. Let $e_a^d$, $e_p^d$, and $e_n^d$ denote the selected codeword embeddings of the anchor, positive, and negative items at the $d$-th quantization layer. We apply a temperature-scaled InfoNCE loss over SID prefixes:
\begin{equation} \mathcal{L}_{\mathrm{cont}} = - \frac{1}{D-1} \sum_{d=1}^{D-1} w_d \log \frac{ \exp(\mathrm{cos}(e_a^d, e_p^d) / \tau) }{ \exp(\mathrm{cos}(e_a^d, e_p^d) / \tau) + \sum_{n} \exp(\mathrm{cos}(e_a^d, e_n^d) / \tau) },
\end{equation}
where $\tau$ is the temperature and $w_d$ controls the importance of each prefix layer. We deliberately exclude the last codebook layer from contrastive alignment. This design encourages high-level SID prefixes to capture shared semantics, while preserving the final layer for fine-grained item discrimination. The overall PCRQ-VAE objective is: 
\begin{equation} \mathcal{L}_{\mathrm{PCRQ\text{-}VAE}} = \mathcal{L}_{\mathrm{recon}} + \lambda_1 \mathcal{L}_{\mathrm{global\_quan}} + \lambda_2 \mathcal{L}_{\mathrm{cont}} . \end{equation}
where $\mathcal{L}_{\mathrm{recon}}$ is the reconstruction loss of standard RQ-VAE, $\lambda_1$ and $\lambda_2$ weight the quantization and contrastive terms. This quantization ensures that similarity has been internalized into the ID prefixes. Consequently, slate recommendation can directly impose constraints and rewards on prefixes to control diversity and relevance during decoding, without resorting to costly a posteriori comparisons over aggregated continuous embeddings.

\subsection{Hierarchical Slate Decoder (HSD)} \label{HSdecoder}
A straightforward generative approach treats each item's SID sequence as tokens and generates the whole slate autoregressively. However, since each item consists of $D$ SIDs, producing an $M$-item slate requires $M \times D$ sequential decoding steps, leading to high latency in real-time recommendation. Moreover, token-level generation entangles intra-item semantic composition with inter-item transition modeling, making it difficult to explicitly capture global slate structure. To address these issues, we propose the \emph{Hierarchical Slate Decoder} (HSD), which decomposes slate generation into a coarse-to-fine process. As shown in Figure \ref{HSD}, HSD contains two components: a coarse-grained slate planner that predicts slate-level preference representations, and a fine-grained item generator that instantiates each preference representation into concrete SIDs. Given user features $u$ and historical behaviors $S^u$, the context encoder first produces a context representation $C = \mathcal{E}(u,S^u)$, which serves as the key and value for cross-attention in both stages.

\subsubsection{Coarse-grained Slate Planner}
The slate planner autoregressively models the global structure of the target slate in the preference embedding space. 
During training, the input sequence is constructed from a beginning-of-sequence token and the preference embeddings of previous items:
\begin{equation}
    h^{(0)} = [\mathrm{Emb}(\mathrm{BOS}), i_1, \cdots, i_{M-1}] \in \mathbb{R}^{M \times d_{\mathrm{model}}},
\end{equation}
where $i_m$ is the preference embedding of item $m$, and $d_\text{model}$ is the hidden dimension.
Following the coarse-to-fine semantics of SIDs, we define it as the sum of the corresponding SID embeddings:
\begin{equation}
    i_m = \sum_{d=1}^{D} s_m^d.
\end{equation}
where $s_m^d$ is the $d$-th SID embedding of item $m$. After $l_{\mathrm{slate}}$ Transformer layers with cross-attention over $C$, the planner outputs the predicted preference sequence:
\begin{equation}
    \hat{I} = \mathrm{Projector}(\mathrm{TransformerDecoder}(C, h^{(0)}))
\end{equation}
where $\hat{I} = [\hat{i}_1, \cdots, \hat{i}_M]$ represents the planned preference embeddings for the slate. $\mathrm{Projector}(\cdot)$ is implemented as a lightweight linear transformation that maps the preference embedding into the SID embedding space for subsequent fine-grained item generation.

\subsubsection{Fine-grained Item Generator}
The item generator grounds each planned preference embedding into a concrete SID sequence. 
For the $m$-th item, given the planned preference embedding $\hat{i}_m$, the input sequence is constructed by prepending $\hat{i}_m$ to the previous ground-truth SIDs during training:
\begin{equation}
    g_m^{(0)} = [\hat{i}_m, s_m^1, \cdots, s_m^{D-1}] \in \mathbb{R}^{D \times d_{\mathrm{model}}},
\end{equation}
where $s_m^d$ denotes the $d$-th SID embedding of the $m$-th item. 
After $l_{\mathrm{item}}$ Transformer layers with cross-attention over $C$, the item generator predicts the SID sequence as:
\begin{equation}
    \hat{S}_m = \mathrm{Linear}(\mathrm{TransformerDecoder}(C, g_m^{(0)})),
\end{equation}
where $\hat{S}_m = [\hat{s}_m^1, \cdots, \hat{s}_m^D]$ denotes the predicted SID logits of the $m$-th item. 
The same fine-grained item generator is shared across all slate positions, which improves parameter efficiency and supports variable-length slate generation.




\subsubsection{Controllable Preference Planning.}
A key advantage of HSD is that the slate-level preference embeddings provide an explicit interface for controlling global slate properties before item decoding. Besides the token-level slate generation objective,
\begin{equation}
    \mathcal{L}_\text{slate}
    =
    -\sum_{m=1}^{M}\sum_{d=1}^{D}
    \log P(\hat{s}_m^{d+1}|
    \hat{i}_1,\cdots,\hat{i}_{m-1},
    \hat{s}_m^{1},\cdots,\hat{s}_m^{d};\Theta),
\end{equation}
where $\Theta$ denotes the model parameters, we can further regularize the geometry of planned preference embeddings to encourage desirable list-level structures. In this work, we introduce an intra-list diversity (ILD) term to reduce redundancy among planned items:
\begin{equation}
    \mathcal{L}_{ILD}
    =
    -\frac{1}{M(M-1)}
    \sum_{m=1}^{M}\sum_{n\neq m}
    (1-\cos(\hat{i}_m, \hat{i}_n)).
\end{equation}
By minimizing pairwise similarity among preference embeddings, HSD encourages the item generator to produce a more diverse slate without expensive post-hoc diversity estimation over decoded items. The final training objective is:
\begin{equation}
    \mathcal{L}_{HSD}
    =
    \mathcal{L}_\text{slate}
    +
    \beta \mathcal{L}_{ILD},
\end{equation}
where $\beta$ controls the strength of diversity regularization. This design preserves accurate SID generation while enabling flexible control over slate-level diversity.


\subsubsection{Efficient Coarse-to-Fine Inference.}
HSD accelerates slate generation by shifting global autoregressive modeling from fine-grained SID tokens to coarse-grained preference embeddings. Instead of applying beam search over the full $M \times D$ SID sequence, the slate planner first generates $M$ preference embeddings to capture global slate intent. Then, instead of continuing beam search across the entire slate, a lightweight item generator performs beam search locally to decode the $D$-token SID sequence conditioned on each preference embedding. This design restricts the time-consuming beam search to short item-level decoding while preserving global slate planning at the preference-embedding level. As a result, HSD significantly reduces the effective search space and inference latency while maintaining high-quality slate generation.

\subsection{Multi-objective Preference Alignment} \label{Training}
Recently, RL-based post-training has been widely adopted to enhance model performance. However, existing approaches overemphasize item-level preference alignment, leaving them susceptible to historical exposure bias and unable to capture the holistic and comparative nature of user slate evaluation. To enhance slate generation quality, we propose a listwise preference alignment mechanism that jointly optimizes three core objectives: ranking fidelity (O1), genuine user interest (O2), and slate diversity (O3).


We adopt Odds Ratio Preference Optimization (ORPO) \cite{hong2024orpo} for efficient and stable alignment. As a reference-free method, ORPO reduces forward computation and GPU memory cost, enabling high-throughput post-training. Moreover, by combining preference optimization with supervised learning, it mitigates objective drift and preserves the pretrained semantic distribution. 
We further construct preference pairs from implicit slate-level feedback, such as watch time, completion rate, and user satisfaction, so that HiGR is directly aligned with key online business metrics.

\subsubsection{Construction of Slate-Level Preference pairs.} 
We construct slate-level preference pairs $(y^+,y^-)$ to support ``triple-objective'' optimization.
Positive slates ($y^+$): Rerank a user's engaged watch sequence in descending order based on true user feedback.
Negative slates ($y^-$) are designed according to our three optimization objectives, respectively:
\begin{itemize}[leftmargin=*]
  \item Randomly permute $y^{+}$ to disrupt  its optimal ordering;
  \item Replace items in $y$ with those receiving negative feedback to filter exposure noise;
  \item Anchor the first item in $y+$ and append top-$(M-1)$ semantically similar items to discourage repetitive recommendations and ``information cocoons'', thereby promoting slate diversity.
\end{itemize}
By performing contrastive optimization between $y^+$ and these diverse negatives, our HiGR model learns to reject suboptimal slates that are poorly ranked, irrelevant, or repetitive.

\begin{table*}[t]
    \centering
    \caption{Offline Performance of HiGR and baselines on industrial and public datasets.}
    \vspace{-0.2cm}
    \label{Overall_exp}
    \setlength{\tabcolsep}{6pt} 
    \begin{tabular}{l|ccccc|ccccc}
        \toprule
        \multirow{3}{*}{\textbf{Methods}} & \multicolumn{5}{c|}{\textbf{Our industrial dataset}} & \multicolumn{5}{c}{\textbf{KuaiRec}} \\
        \cmidrule(lr){2-6} \cmidrule(lr){7-11}
        & \multicolumn{2}{c}{Impressions} & \multicolumn{2}{c}{Effective Views} & \multirow{2}{*}{NDCG@5} & \multicolumn{2}{c}{Impressions} & \multicolumn{2}{c}{Effective Views} & \multirow{2}{*}{NDCG@5} \\
        \cmidrule(lr){2-3} \cmidrule(lr){4-5} \cmidrule(lr){7-8} \cmidrule(lr){9-10}
        & hit@5 & recall@5 & hit@5 & recall@5 & & hit@5 & recall@5 & hit@5 & recall@5 & \\
        \midrule
        ListCVAE 	& 0.0857 & 0.0178 & 0.0571 & 0.0117 & 0.0186 & 0.2304 & 0.0523 & 0.0757 & 0.0161 & 0.0630 \\
        \midrule
        BertRec 	& 0.0911 & 0.0187 & 0.0632 & 0.0129 & 0.0201 & 0.2612 & 0.0583 & 0.0829 & 0.0177 & 0.0713 \\
        SASRec 	    & 0.1057 & 0.0218 & 0.0700 & 0.0143 & 0.0243 & 0.3017 & 0.0756 & 0.1028 & 0.0231 & 0.0839 \\
        \midrule
        TIGER 	& 0.1812 & 0.0383 & 0.1204 & 0.0249 & 0.0406 & 0.4455 & 0.1299 & 0.1566 & 0.0378 & 0.1363 \\
        HSTU 	& 0.2281 & 0.0506 & 0.1487 & 0.0310 & 0.0492 & 0.5055 & 0.1334 & 0.1728 & 0.0397 & 0.1466 \\
        OneRec-25M 	& 0.2438 & 0.0577 & 0.1603 & 0.0367 & 0.0589 & 0.5150 & 0.1395 & 0.1782 & 0.0422 & 0.1496 \\
        HiGR-25M-w/o RL 	& 0.2641 & 0.0612 & 0.1810 & 0.0395 & 0.0631 & 0.5218 & 0.1437 & 0.1805 & 0.0462 & 0.1536 \\
        \textbf{HiGR-25M} 	& \textbf{0.2825} & \textbf{0.0692} & \textbf{0.1945} & \textbf{0.0433} & \textbf{0.0714} & \textbf{0.5290} & \textbf{0.1485} & \textbf{0.1846} & \textbf{0.0491} & \textbf{0.1574} \\
        \textbf{HiGR-100M} 	& \textbf{0.3163} & \textbf{0.0760} & \textbf{0.2145} & \textbf{0.0495} & \textbf{0.0831} & \textbf{0.5360} & \textbf{0.1544} & \textbf{0.1905} & \textbf{0.0546} & \textbf{0.1651} \\
        \bottomrule
    \end{tabular}
    \vspace{-0.2cm}
\end{table*}

\subsubsection{ORPO-based Post-Training Loss Function.}
We define the per-step log-odds for a slate $y$ as,
\begin{equation}
    \ell_\theta(x,y_t) = \log\frac{\pi_{\theta}(y_{t}\mid x,y_{<t})}{1-\pi_{\theta}(y_{t}\mid x,y_{<t})}
\end{equation}
where $\pi_\theta$ is the policy actor.

Let $z_\theta=\sum_{t=1}^{MD}\ell_\theta(x,y_t)$ denote the slate-level log-odds. For preference pair $\{(y^{+},y^{-})\}$, the combined post-training objective is:
\begin{equation}
    \mathcal{L}_\mathrm{post} = - \log\pi_\theta(y_t^{+}|x,y_{i<t}^{+}) -\alpha \log\sigma\left(z_\theta(x,y_t^{+}) - z_\theta(x,y_t^{-})\right)
\end{equation}
where $\alpha$ is the preference alignment coefficient.
This combined loss enables the model to learn true user preferences while preserving generative accuracy without an auxiliary reference model.
In summary, our post-training approach transforms preference alignment into a holistic slate optimization process, ensuring generated SID sequences are accurate, diverse and well-ordered.

\section{System Deployment}
Training follows a three-stage pipeline (PCRQ-VAE, HSD pretraining, ORPO alignment) on filtered high-quality slate-level samples. In online deployment, HiGR is deployed on clusters with tens of NVIDIA H20 and L20 GPUs for training and serving, respectively. For each request, the online feature assembler retrieves real-time user features, and HiGR generates slate SIDs via HSD coarse-to-fine decoding with KV caching. The SIDs are mapped back to items through the SID-to-item index, then filtered by deduplication, availability, safety, and business rules before being returned to users, while exposure and feedback logs are sent back for continuous updates. At the same throughput, HiGR keeps P99 latency below 50ms and reduces GPU demand by about 60\% over OneRec, showing strong cost-effectiveness for industrial slate recommendation.

\section{Experiments}
To verify the effectiveness of HiGR, we have conducted comprehensive experiments to solve the following five key research questions:
\begin{itemize}[leftmargin=*]
    \item \textbf{RQ1:} How does HiGR compare against SOTA baseline models?
    \item \textbf{RQ2:} What is the contribution of each core component in HiGR?
    \item \textbf{RQ3:} Does the performance of HiGR follow scaling laws?
    \item \textbf{RQ4:} How efficient is HiGR?
    \item \textbf{RQ5:} How does HiGR perform in real-world deployments?
\end{itemize}

\subsection{Experimental Setup}
\subsubsection{Baseline Methods}
For evaluation, we compare HiGR with representative baselines across three pretraining paradigms: traditional slate recommendation, \textbf{ListCVAE} \cite{jiang2018beyond}; discriminative models, \textbf{SASRec} \cite{kang2018self} and \textbf{BERT4Rec} \cite{sun2019bert4rec}; and recent generative methods, \textbf{TIGER} \cite{rajput2023recommender}, \textbf{HSTU} \cite{zhai2024actions}, and \textbf{OneRec} \cite{deng2025onerec}. For post-training, we further compare with \textbf{DPO} \cite{rafailov2024directpreferenceoptimizationlanguage}.

\subsubsection{Datasets and Metrics}
We evaluate HiGR on a large-scale Tencent commercial media platform and the public KuaiRec dataset~\cite{Gao_2022} with offline metrics and online A/B tests. 
For industrial dataset, we used 1 billion samples from all users for pre-training, selected 3\% of samples for post-training. For KuaiRec, we build chronological user sequences and adopt leave-five-out slate evaluation. 
Offline metrics include Effective View-based Hit Rate, Recall, and NDCG on positive-feedback slates, as well as Impression-based Hit Rate and Recall from exposure logs to assess exposure-pattern modeling.

\subsubsection{Implementation Details}
We set $d_\text{model}=512$, $d_\text{FFN}=2048$, $D=3$, and the codebook size to $1024$. For PCRQ-VAE, we set $\eta=0.1$, $\lambda_1=0.1$, and $\lambda_2=0.01$ in Eq.~(4), and $w_1 = 1$, $w_2 = 0.1$ in Eq.~(3). For preference alignment, we set the coefficient $\alpha=0.1$ in Eq.~(14).


\begin{table}
    \caption{Performance comparison of baselines and our PCRQ-VAE with different prefix contrastive constraints.}
    \vspace{-0.2cm}
    \label{CSID_result}
    \resizebox{\linewidth}{!}{
    \begin{tabular}{l|ccccc}
        \toprule
        \multirow{2}{*}{Methods} & \multirow{2}{*}{RQ-Kmeans} & \multirow{2}{*}{RQ-VAE} & \multicolumn{3}{c}{PCRQ-VAE} \\
        \cmidrule(lr){4-6}
        & & & prefix1 & prefix2 & prefix3 \\
        \midrule
        Collision $\downarrow$ & 0.2126 & 0.0298 & 0.0264 & 0.0237 & 0.0873 \\
        Concentration $\uparrow$ & 0.83 & 0.75 & 0.88 & 0.93 & 0.96 \\
        Consistency $\uparrow$ & 0.4533 & 0.5577 & 0.6205 & 0.6647 & 0.6932 \\
        \bottomrule
    \end{tabular}
    }
    \vspace{-0.5cm}
\end{table}

\subsection{Main Results (RQ1)}

Table \ref{Overall_exp} presents the overall comparison results between HiGR and other representative recommendation methods. HiGR consistently outperforms all methods on both the industrial dataset and KuaiRec across all metrics, demonstrating strong effectiveness and robustness. The gains over traditional slate recommendation, discriminative models, and recent generative baselines validate the benefits of prefix-contrastive RQ-VAE, hierarchical slate generation and listwise preference alignment.

\subsection{Ablation Study (RQ2)}
\subsubsection{Performance of PCRQ-VAE}
Table \ref{CSID_result} evaluates PCRQ-VAE against baselines and ablates our prefix-level constraint design. We use three metrics for SID quality. \textbf{Collision}: The ratio of distinct items mapped to the same SID; \textbf{Concentration}: The category purity of items sharing the same prefix; \textbf{Consistency}: The pairwise category agreement within each prefix group. Our optimal configuration (\textbf{prefix2-contra}), which applies contrastive constraints strictly to the first $D-1$ layers, remarkably reduces the collision rate to 2.37\% while achieving 93\% concentration and 66.47\% consistency (outperforming RQ-VAE by $\sim$10.7\%). This shows that contrastive learning improves the codebook space, while the $D-1$ constraint preserves early-layer semantic hierarchy and leaves the final layer for unique item identities. Extending constraints to the leaf level (\textbf{prefix3-contra}) risks "identity collapse"; it yields only marginal consistency gains but inflates collision nearly 4$\times$ (0.0237 → 0.0873). Thus, prefix-level constraints are essential to achieve semantic consistency without compromising fine-grained item discriminability.

\subsubsection{Ablation Study of HSD Component}

\begin{table}
    \caption{Ablation experiments of HSD variants.}
    \vspace{-0.2cm}
    \label{HSG_result}
    \resizebox{\linewidth}{!}{
    \begin{tabular}{l|cc|cc|c|c}
        \toprule
        \multirow{2}{*}{Variant} & \multicolumn{2}{c|}{Impressions} &  \multicolumn{2}{c|}{Effective Views} & \multirow{2}{*}{NDCG@5} & \multirow{2}{*}{ILD}\\
         & hit@5 & recall@5 & hit@5 & recall@5 && \\
        \midrule
        \textbf{HiGR} 	    & \textbf{0.2994} & \textbf{0.0721} & \textbf{0.2012} & \textbf{0.0465} & \textbf{0.0753}&\textbf{0.62} \\
        \hspace{0.5em} - w/o $\mathcal{L}_{ILD}$ 	& 0.2997 & 0.0722 & 0.2009 & 0.0464 & 0.0755&0.59 \\
        \hspace{0.5em} - w/o CE 	& 0.2785 & 0.0651 & 0.1837 & 0.0418 & 0.0664&- \\
        \hspace{0.5em} - Non-shared & 0.2973 & 0.0735 & 0.2002 & 0.0477 & 0.0748&- \\
        \bottomrule
    \end{tabular}
    }
    \vspace{-0.5cm}
\end{table}

Table \ref{HSG_result} studies three HSD designs: diversity regularization, context embedding, and item-generator sharing. Removing $\mathcal{L}_{ILD}$ keeps accuracy metrics nearly unchanged but reduces ILD from 0.62 to 0.59, showing its effectiveness in improving intra-list diversity.  Removing context embedding causes a notable performance drop, confirming its importance for conditional SID decoding. In addition, non-shared item generators bring no clear gain, we therefore employ a shared item generator to ensure efficiency and handle variable-length sequences.

\subsubsection{Ablation Study of Preference Alignment}

\begin{table}
    \caption{Intra-list Diversity (ILD) of HiGR and baselines on industrial and public datasets.}
    \vspace{-0.4cm}
    \label{ablation_rl}
    \resizebox{\linewidth}{!}{
    \begin{tabular}{l|ccc|ccc}
        \toprule
        \multirow{2}{*}{Datasets} & \multirow{2}{*}{TIGER} & \multirow{2}{*}{HSTU} & \multirow{2}{*}{OneRec} & \multicolumn{3}{c}{HiGR} \\
        \cmidrule(lr){5-7}
        & & & & w/o RL & DPO & ORPO\\
        \midrule
        Industrial & 0.38 & 0.42 & 0.48 & 0.52 & 0.57 & 0.62 \\
        KuaiRec & 0.40 & 0.43 & 0.50 & 0.55 & 0.60 & 0.63 \\
        \bottomrule
    \end{tabular}
    }
    \vspace{-0.3cm}
\end{table}
Table \ref{ablation_rl} reports the Intra-list Diversity (ILD) of various models on two datasets. Even without reinforcement learning (RL), our HiGR model outperforms baselines (TIGER, HSTU, OneRec). Moreover, RL-based post-training further enhances diversity, with ORPO achieving the best overall ILD scores (0.62 and 0.63), demonstrating its superiority over DPO.

\subsubsection{Effect of Layer Configurations}
We evaluate different layer allocations between the slate planner and item generator under a fixed total depth. 
Since slate planning involves global list-level modeling, whereas item generation is a lightweight SID decoding task, assigning more layers to the planner is more effective. As shown in Table \ref{ablation_layer}, the default $14{:}2$ configuration provides the best quality-latency trade-off. Compared with $12{:}4$, it reduces inference latency by $26\%$ while causing only a $0.6\%$ drop in NDCG@5. Thus, we use $l_{\mathrm{slate}}=14$ and $l_{\mathrm{item}}=2$ by default.

\begin{table}[htbp]
    \centering
    \caption{Ablation study of layer configurations on recommendation quality and inference latency.}
    \vspace{-0.4cm}
    \label{ablation_layer}
    \begin{tabular}{l|ccccc} 
        \toprule
        $L_{\text{slate}}$ + $L_{\text{item}}$  & 15 + 1 & 14 + 2 & 12 + 4 & 10 + 6 & 8 + 8 \\
        \midrule
        NDCG@5 & 0.0623 & 0.0631 & 0.0635 & 0.0621 & 0.0605 \\
        Relative Latency & 0.932$\times$ & 1.000$\times$ & 1.355$\times$ & 1.710$\times$ & 2.065$\times$ \\
        \bottomrule
    \end{tabular}
    \vspace{-0.4cm}
\end{table}


\subsubsection{Performance of HiGR with Different Preference Objectives}
\begin{table}[htbp]
    \centering  
    \vspace{-0.5cm}
    \caption{Performance comparison of HiGR with different preference objectives.}
    \vspace{-0.4cm}
    \label{ablation_rl_component}
    
    \setlength{\tabcolsep}{12pt} 
    
    \begin{tabular}{ccc|cc}
        \toprule
        O1 & O2 & O3  & NDCG@5 & ILD\\
        \midrule
        \ding{56} & \ding{56} & \ding{56} & 0.0753 & 0.52\\
        \ding{52} & \ding{56} & \ding{56} & 0.0812 & 0.54\\
        \ding{56} & \ding{52} & \ding{56} & 0.0792 & 0.55\\
        \ding{56} & \ding{56} & \ding{52} & 0.0780 & 0.64\\
        \ding{52} & \ding{52} & \ding{52} & 0.0831 & 0.62\\
        \bottomrule
    \end{tabular}
    \vspace{-0.3cm}
\end{table}

We evaluate the three objectives defined in Section \ref{Training}, i.e., ranking fidelity (O1), genuine interest (O2), and slate diversity (O3). Symbols \ding{52} and \ding{56} denote the inclusion and exclusion of an objective during ORPO training, respectively. As shown in Table \ref{ablation_rl_component}, the three objectives play differentiated roles: O1 and O2 mainly improve the ranking metric NDCG@5 (0.0812 and 0.0792 vs.\ 0.0780 for O3 alone), as both are ranking-oriented signals, while O3 primarily boosts the diversity metric ILD (0.64 vs.\ 0.54/0.55). Jointly optimizing all three achieves the best NDCG@5 (0.0831) without sacrificing ILD (0.62), showing that ranking accuracy and diversity are complementary rather than conflicting.

\subsection{Scaling-Law Validation (RQ3)}

To investigate the scalability of our proposed framework, we evaluated the loss and NDCG@5 of HiGR across a broad spectrum of model sizes, spanning 0.05B to 2B parameters. As shown in Figure \ref{scale}, the results exhibit a clear linear trajectory on a logarithmic scale, revealing a power-law relationship between model capacity and performance. This property is critical for industrial applications, as it implies that predictable performance gains can be consistently achieved by scaling model capacity, thereby offering a reliable pathway to improve recommendation quality in large-scale scenarios.

\subsection{Efficiency Analysis (RQ4)} 
\begin{figure}[htbp] 
    \vspace{-0.5cm}
  \centering
  \begin{minipage}[b]{0.48\linewidth} 
    \centering
    \includegraphics[width=\linewidth]{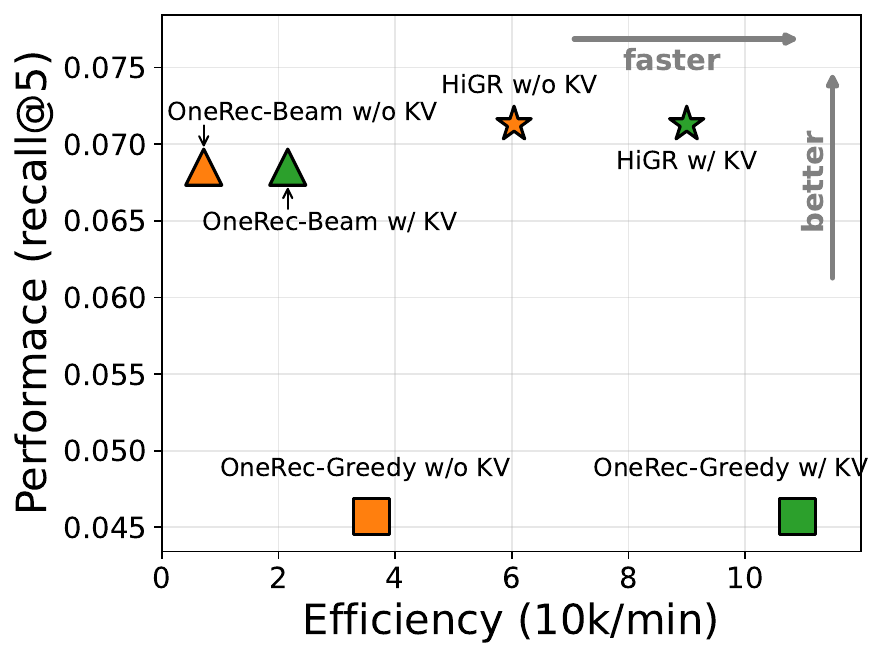}
    \caption{Efficiency comparison of HiGR and OneRec.}
    \label{efficiency}
  \end{minipage}
  \hfill
  \begin{minipage}[b]{0.51\linewidth} 
    \centering
    \includegraphics[width=\linewidth]{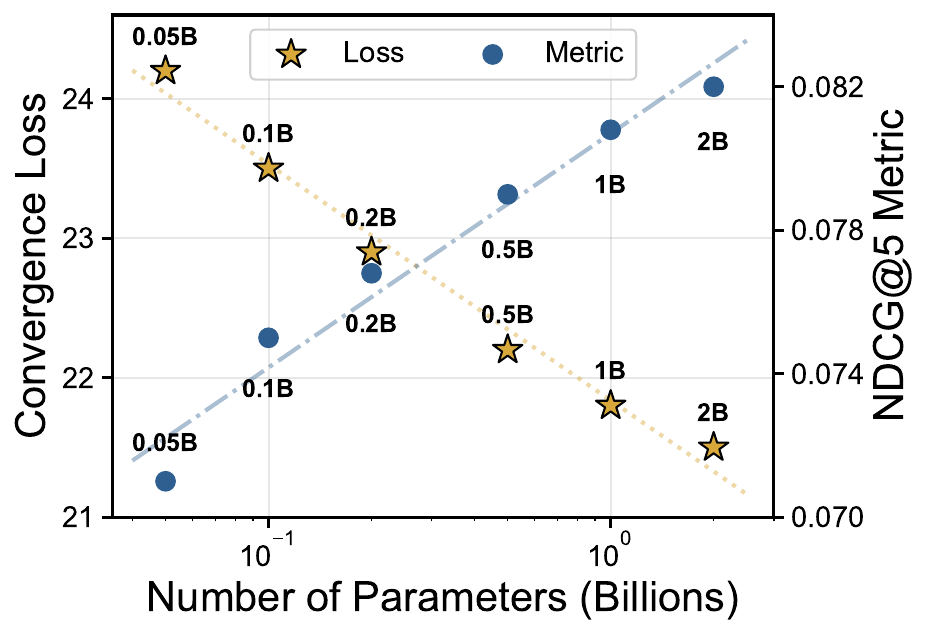}
    \caption{Scaling trends of HiGR on loss and NDCG@5.}
    \label{scale}
  \end{minipage}
  \vspace{-0.7cm}
\end{figure}
We validate HiGR's efficiency through complexity analysis and empirical evaluations. For a slate of $M$ items with $D$ SIDs each, decoding involves $M \times D$ tokens. Let $d$ be the hidden dimension, $l_\text{slate}$ the number of slate-planners in HiGR and lazy decoders in OneRec, and $l_\text{item}$ the number of item-generators. Decomposing decoding into greedy slate planning with beam SID generation drops complexity from $\mathcal{O}(B \cdot M^3D^3\cdot l_\text{slate} \cdot d)$ (OneRec) to $\mathcal{O}(M^3 \cdot l_\text{slate} \cdot d + B \cdot MD^3 \cdot l_\text{item} \cdot d)$, theoretically demonstrating that HiGR has substantially lower decoding complexity than OneRec. For a fair empirical comparison, OneRec and HiGR use identical model settings and the same hardware environment. As shown in figure \ref{efficiency}, without KV caching, HiGR achieves over $5\%$ higher Recall@5 than OneRec-Beam with more than $5\times$ speedup, and also surpasses OneRec-Greedy in both accuracy and speed. With KV caching, it still offers the best quality-efficiency trade-off.

\subsection{Online A/B Tests (RQ5)}
We conducted online A/B tests with 5\% live traffic on Tencent's commercial media platform serving hundreds of millions of users. Compared with the incumbent multi-stage system, HiGR improves \textbf{Average Stay Time}, \textbf{Average Watch Time}, \textbf{Average Video Views}, and \textbf{Average Request Count} by \textbf{1.03\%}, \textbf{1.22\%}, \textbf{1.73\%}, and \textbf{1.57\%}, respectively, demonstrating clear real-world engagement gains.

\section{Conclusion}
We propose HiGR, a hierarchical generative framework for efficient slate recommendation. 
HiGR combines Prefix-Contrastive RQ-VAE for structured item tokenization, Hierarchical Slate Decoder for coarse-to-fine slate generation, and listwise preference alignment for holistic slate optimization. 
Extensive offline and online results show that HiGR improves both recommendation quality and inference efficiency, demonstrating the potential of hierarchical generative modeling for industrial slate recommendation.

\clearpage
\section*{Generative AI Usage Disclosure}
We acknowledge the use of generative AI (e.g., ChatGPT) strictly for manuscript copyediting (grammar and phrasing) and software engineering assistance (code snippets and debugging). All scientific contributions, including idea conceptualization, methodology design, and data analysis, are entirely human-authored.
\bibliographystyle{ACM-Reference-Format}
\bibliography{ref-base}

\end{document}